\documentclass[reprint,aps,prc,twocolumn,superscriptaddress,floatfix,10pt]{revtex4-2}
\usepackage[utf8]{inputenc}
\usepackage{bm}
\usepackage{times}
\usepackage{amssymb,amsbsy,amsmath,amsfonts}
\usepackage{graphicx}
\usepackage{float}
\usepackage{color}
\usepackage{xcolor}
\usepackage[colorlinks,allcolors=blue]{hyperref}
\makeatletter
\def\NAT@def@citea{\def\@citea{\NAT@separator}}
\makeatother

\begin{document}

\title{Evolution of $N=20,28,50$ shell closures in the $ 20 \leqslant Z \leqslant 30$  region in deformed relativistic Hartree-Bogoliubov theory in continuum}

\author{Ru-You Zheng}
\affiliation{School of Physics, Beihang University, Beijing, 102206, China}
\author{Xiang-Xiang Sun}
\affiliation{School of Nuclear Science and Technology,
 University of Chinese Academy of Sciences,
 Beijing 100049, China}
 \affiliation{Institut für Kernphysik, Institute for Advanced Simulation and Jülich Center for Hadron Physics, Forschungszentrum Jülich, D-52425 Jülich, Germany}
 \affiliation{CAS Key Laboratory of Theoretical Physics,
              Institute of Theoretical Physics,
              Chinese Academy of Sciences,
              Beijing 100190, China}
\author{Guo-fang Shen}
\affiliation{School of Physics, Beihang University, Beijing, 102206, China}

\author{Li-Sheng Geng}
\email[Corresponding author: ]{lisheng.geng@buaa.edu.cn}

\affiliation{School of Physics, Beihang University, Beijing, 102206, China}
\affiliation{Peng Huanwu Collaborative Center for Research and Education, Beihang University, Beijing 100191, China}
\affiliation{Beijing Key Laboratory of Advanced Nuclear Materials and Physics, Beihang University, Beijing, 102206, China}
\affiliation{Southern Center for Nuclear-Science Theory (SCNT), Institute of Modern Physics, Chinese Academy of Sciences, Huizhou 516000, Guangdong Province, China}

\begin{abstract}
Magicity, or shell closure, plays an important role in our understanding of complex nuclear phenomena. In this work, we employ one of the state-of-the-art density functional theories, the deformed relativistic Hartree-Bogoliubov theory in continuum (DRHBc) with the density functional PC-PK1, to investigate the evolution of the $N=20,28,50$ shell closures in
the $ 20 \leqslant Z \leqslant 30$ region. We show how these three conventional shell closures evolve from the proton drip line to the neutron drip line by studying the charge radii, two-neutron separation energies, two-neutron gaps, quadrupole deformations, and single-particle levels. In particular, we find that in the $ 21 \leqslant Z \leqslant 27$ region, the $N=50$ shell closure disappears or becomes quenched,  mainly due to the deformation effects. Similarly, both experimental data and theoretical predictions indicate that the $N=28$ shell closure disappears in the Mn isotopic chain, also predominantly due to the deformation effects. The DRHBc theory predicts the existence of the $N=20$ shell closure in the Ca, Sc, and Ti isotopic chains, but the existing data for the Ti isotopes suggests the contrary, and therefore more investigations are needed. 
\end{abstract}

\maketitle

\section{Introduction}
The shell structure of atomic nuclei, in particular, shell closure or magicity,  plays an important role in nuclear physics and nuclear astrophysics~\cite{Otsuka:2018bqq}. Worldwide, many Radioactive Ion Beam (RIB) facilities have been built or are under construction, which enable us to study the shell structure of atomic nuclei and to explore the limits of their existence,  e.g., the RIB Factory (RIBF) at RIKEN in Japan~\cite{Kamigaito:2020pll}, the Facility for Rare Isotope Beams (FRIB) in the United States of America~\cite{Castelvecchi2022}, the High Intensity heavy-ion Accelerator Facility (HIAF) in China ~\cite{Zhou2022StatusOT}, the Facility for Antiproton and Ion Research (FAIR) in Germany~\cite{Sturm:2010yit}, and the Rare isotope Accelerator complex for ON-line
experiments (RAON) in Korea~\cite{Sohn:2022veb}. The operation of these large scientific installations will certainly advance our understanding of the strong nuclear force and the atomic nuclei from which the visible universe is formed.

These RIB facilities have made many exciting discoveries, such as the occurrence of new shell gaps, which results in new magic numbers. For example, the studies of $^{52} \mathrm{Ca}$~\cite{Gade:2006dp,Wienholtz:2013nya}, $^{54} \mathrm{Ti}$~\cite{Dinca:2005wm}, and $^{56} \mathrm{Cr}$~\cite{RISING:2005nzn} provide substantial evidence for the onset of a shell closure at $N=32$. In Ref.~\cite{Steppenbeck:2013mga}, direct experimental evidence for a new  magic number of  $N=34$ is found in the neutron-rich calcium isotopes, and the shell closures at $N=32$ and $N=34$ are shown to be driven by the tensor force. However, the signature  of the $N =34$ shell closure was not confirmed by the two-neutron shell gaps of Ti and V isotopes~\cite{Iimura:2022fdg}. In recent years, there have been extensive studies on whether the $N = 40$ subshell closure is a local phenomenon that only exists in the magic nickel chain~\cite{Malbrunot-Ettenauer:2021fnr,deGroote:2019yqm,Babcock:2016rkv}. It is found that the charge radii of copper isotopes only reveal a weak $N=40$ subshell closure effect~\cite{Bissell:2016vgn}, and the chromium isotopes form a new island of inversion at $N=40$~\cite{Mougeot:2018ttn}. For the traditional magic numbers of 28 and 50, Ref.~\cite{Taniuchi:2019pen} provided the first direct experimental evidence for the doubly-magic nature of ${ }^{78} \mathrm{Ni}$, and this experiment also confirmed the existence of a deformed second low-energy $2^{+}$ state, supporting the prediction of shape coexistence in ${ }^{78} \mathrm{Ni}$~\cite{Nowacki:2016isq}. The first measurement of the charge radius of ${ }^{56} \mathrm{Ni}$ provides direct support for its doubly magic nature~\cite{Sommer:2022sok}. 

Meanwhile, tremendous theoretical efforts have been made to understand shell evolution. Among them, covariant density functional theories (CDFTs) have received a lot of attention due to their successful descriptions of various nuclear phenomena throughout the nuclear chart~\cite{Meng:2005jv,Liang:2014dma,Meng2016book,Shen:2019dls,Yang:2019fvs}. 
To accurately describe exotic nuclei close to drip line, it is essential to consider the pairing correlations and couplings to the continuum~\cite{Meng:2002ps,XIA20181}. Additionally, it should be noted that most open-shell nuclei are deformed. Therefore, in Refs.~\cite{Zhou:2009sp,Li:2012gv}, the deformed relativistic Hartree-Bogoliubov theory in continuum (DRHBc) was developed and it can self-consistently treat the deformation effects and pairing-induced continuum. Lately, this theory has been applied to describe or predict the ground state properties of deformed halo nuclei~\cite{Zhou:2009sp, Li:2012xaa, Sun:2018ekv,Zhang:2019qeu, Sun:2020tas, Yang:2021pbl, Sun:2021alk, Zhong:2021yhm, Zhang:2023dhj,Zhang:2023bqg}.  In Ref.~\cite{DRHBcMassTable:2020cfw}, the DRHBc theory with the point-coupling density functionals was developed for even-even nuclei. It has recently been applied to construct a mass table for even-even nuclei~\cite{DRHBcMassTable:2022uhi}, and the mass table for odd-$A$ and odd-odd nuclei is under construction~\cite{DRHBcMassTable:2022rvn}. In addition, many interesting studies have been performed, such as the impact of deformation effects on the location of the neutron drip line~\cite{In:2020asf}, the dynamical correlation energy with a two-dimensional collective Hamiltonian~\cite{Sun:2022qck}, the multipole expansion of densities~\cite{Pan:2019gyo}, the rotational mode of deformed halo nuclei~\cite{Sun:2021nfb, Sun:2021nyl}, the bubble structure and shape coexistence~\cite{CHOI:2022bld,Kim:2021skf}, the peninsulas of stability beyond the two-neutron drip line~\cite{Zhang:2021ize,Pan:2021oyq,He:2021thz}, the optimization of the Dirac Woods-Saxon basis~\cite{Zhang:2022yru}, the shell closure  at $N = 82$  in the neodymium isotopic chain~\cite{DRHBcMassTable:2020cfw, DRHBcMassTable:2022rvn}, the collapse of the  $N=28$ shell closure in the newly discovered ${ }^{39} \rm{Na}$~\cite{Zhang:2023dhj}, the odd-even staggering and kink structures of charge radii of $\rm{Hg}$ isotopes~\cite{Mun:2023lfc}, the prolate-shape dominance in atomic nuclei ~\cite{Guo:2023ucm}, the nuclear charge radii and shape evolution of Kr and Sr isotopes~\cite{Zhang:2023fym}, and the one-proton emission of $^{148-151}$Lu using the DRHBc+WKB approach~\cite{Xiao:2023uld}. In this work, we apply the DRHBc theory to study the $ 20 \leqslant Z \leqslant 30$ isotopes. In particular, we focus on the evolution of the $N=20,28,50$ shell closures in this region.

This paper is organized as follows. In Sec.\ref{DRHBc framework}, we briefly introduce the DRHBc theory. Results and discussions are presented in
Sec.\ref{Results and Discussion}, followed by a short summary in Sec.\ref{Summary and Outlook}.

\section{Deformed relativistic Hartree-Bogoliubov theory in continuum} \label{DRHBc framework}
Detailed accounts of the DRHBc theory can be found in Refs.~\cite{DRHBcMassTable:2020cfw,DRHBcMassTable:2022rvn,Li:2012gv,Li:2012xaa}. Here we briefly introduce the formalism for the convenience of discussions. 
In the DRHBc theory, the relativistic Hartree-Bogoliubov (RHB) equation reads,
\begin{align}\label{RHB equation}
\left(\begin{array}{cc}
h_{D}-\lambda_{\tau} & \Delta \\
-\Delta^{*} & -h_{D}^{*}+\lambda_{\tau}
\end{array}\right)\left(\begin{array}{l}
U_{k} \\
V_{k}
\end{array}\right)=E_{k}\left(\begin{array}{c}
U_{k} \\
V_{k}
\end{array}\right),
\end{align}
where $h_{D}$ is the Dirac Hamiltonian, $ \Delta$ is the pairing potential, $\lambda_{\tau}$ is the Fermi energy for neutrons or protons ($\tau=n,p$), $E_{k}$ is the quasiparticle energy, and $U_{k}$ and $V_{k}$ are the quasiparticle wave functions. The Dirac Hamiltonian in coordinate space is
\begin{align}
h_{D}\left(\boldsymbol{r}\right)=\boldsymbol{\alpha} \cdot \boldsymbol{p}+V(\boldsymbol{r})+\beta[M+S(\boldsymbol{r})],
\end{align}
where $M$ is the nucleon mass, and $S(\boldsymbol{r})$ and $V(\boldsymbol{r})$ are the scalar and vector potentials, respectively. The pairing potential reads
\begin{align}\label{pairing potential}
\Delta\left(\boldsymbol{r}_{1}, \boldsymbol{r}_{2}\right)=V^{p p}\left(\boldsymbol{r}_{1}, \boldsymbol{r}_{2}\right) \kappa\left(\boldsymbol{r}_{1}, \boldsymbol{r}_{2}\right),
\end{align}
where $\kappa$ is the pairing tensor~\cite{Ring1980NMBP} and $V^{p p}$ is the pairing force of a density-dependent zero-range type,
\begin{align}
V^{p p}\left(\boldsymbol{r}_{1}, \boldsymbol{r}_{2}\right)=V_{0} \frac{1}{2}\left(1-P^{\sigma}\right) \delta\left(\boldsymbol{r}_{1}-\boldsymbol{r}_{2}\right)\left(1-\frac{\rho\left(\boldsymbol{r}_{1}\right)}{\rho_{\mathrm{sat}}}\right).
\end{align}

For an axially deformed nucleus with spatial reflection symmetry, the potentials and densities can be expanded in terms of Legendre polynomials:
\begin{align}
\label{eq:expansion}
    f(\boldsymbol{r})=\sum_{\lambda} f_{\lambda}(r) P_{\lambda}(\cos \theta), ~\lambda=0,2,4, \cdots.
\end{align}

For an odd-$A$ or odd-odd nucleus, one needs to further take into account the blocking effect of the unpaired nucleon(s)~\cite{Geng:2005yu,Li:2012xaa,Perez-Martin:2008dlm,DRHBcMassTable:2022rvn}. More details about the treatment of blocking effects in the DRHBc theory can be found in Refs. \cite{Li:2012xaa,DRHBcMassTable:2022rvn}.

The RHB equations are solved using the basis expansion method with the Dirac Woods–Saxon (WS) basis~\cite{Zhou:2003jv,Zhou:2009sp,Zhang:2022yru}, which can properly describe the large spatial extension of weakly bound nuclei.
In the numerical calculation, 
the angular momentum cutoff for the Dirac WS basis is chosen to be $J_{\max }=\frac{23}{2} \hbar$.
The maximum expansion order in Eq.~(\ref{eq:expansion}) is $\lambda_{\text{max}}$ = 6, which is sufficient for our study~\cite{WangXiaoQian:2021wym,DRHBcMassTable:2022uhi}.
The size of the box to obtain the WS basis is taken to be 20 fm, and the energy cutoff for the Dirac WS basis in the Fermi sea is $E_{\text {cut }}^{+}=300 ~\mathrm{MeV}$. For the particle-particle channel we use the zero-range pairing force with a saturation density $\rho_{\mathrm{sat}}=0.152 ~\mathrm{fm}^{-3}$ and a pairing strength $V_0=-325\ \mathrm{MeV} \cdot \mathrm{fm}^3$~\cite{DRHBcMassTable:2020cfw,DRHBcMassTable:2022rvn}. 
All the numerical details are the same as those adopted in constructing the DRHBc mass tables~\cite{DRHBcMassTable:2020cfw,DRHBcMassTable:2022rvn}. 

\section{Results and Discussions}\label{Results and Discussion}

To understand the evolution of the $N=20,28$, and $50$ shell closures, we study the charge radii, two-neutron separation energies, two-neutron gaps, quadrupole deformations, and single-particle levels of $20\le Z \le 30$ isotopes in detail.  Based on the systematic calculations using the DRHBc theory, in the following, we show these bulk properties of all the isotopes of $20\le Z \le 30$  from the proton drip line to the neutron drip line. 
Here, we consider a nucleus as
bound only if both the one- and two-nucleon separation energies of this nucleus are positive, which is the same as
the strategy adopted in Ref.~\cite{XIA20181}. The resulting neutron drip line can reach up to $^{80}\rm{Ca}$, $^{83}\rm{Sc}$, $^{84}\rm{Ti}$,  $^{87}\rm{V}$, $^{90}\rm{Cr}$, $^{95}\rm{Mn}$, $^{96}\rm{Fe}$, $^{97}\rm{Co}$, $^{98}\rm{Ni}$, $^{107}\rm{Cu}$, $^{110}\rm{Zn}$, and the positions of the proton drip line are $^{34}\rm{Ca}$, $^{40}\rm{Sc}$, $^{40}\rm{Ti}$,  $^{43}\rm{V}$, $^{43}\rm{Cr}$, $^{46}\rm{Mn}$, $^{47}\rm{Fe}$, $^{49}\rm{Co}$, $^{50}\rm{Ni}$, $^{55}\rm{Cu}$, $^{56}\rm{Zn}$  for each isotopic chain.

\subsection{Charge Radii}\label{charge radii}

\begin{figure}[!htbp]
    \centering
    \includegraphics[width=.5\textwidth]{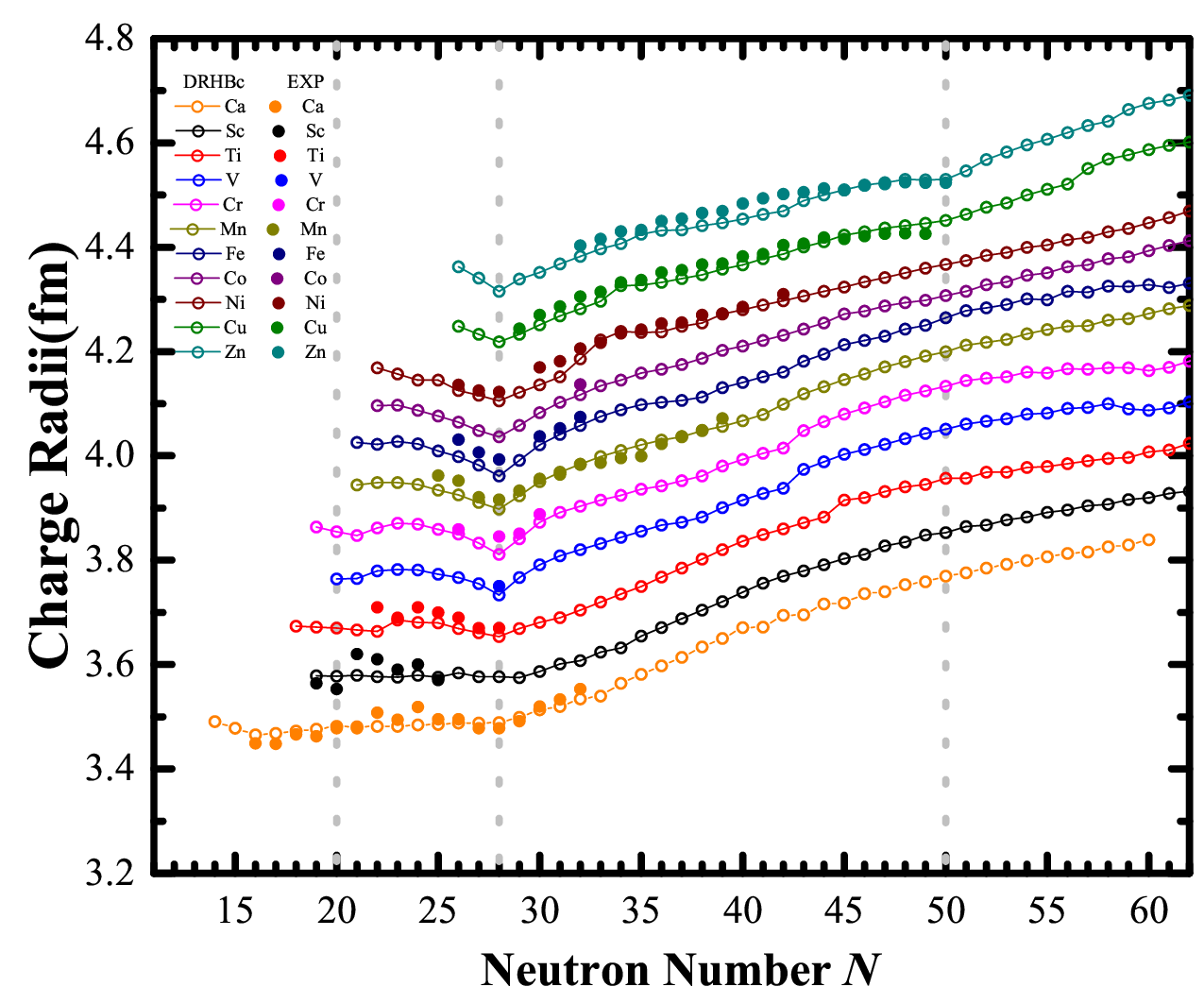}
    \caption{Theoretical charge radii as a function of the neutron number for Ca, Sc, Ti, V, Cr, Mn, Fe, Co, Ni, Cu and Zn isotopes (open circles), where the results for the even-even nuclei are taken from Ref.~\cite{DRHBcMassTable:2022uhi}. The available data from Refs.~\cite{ANGELI201369,Li:2021fmk,Sommer:2022sok,PhysRevLett.131.102501} are also shown for comparison (solid points). In order to visualize the shell closures, 0.05, 0.10, 0.15, 0.20, 0.25, 0.30, 0.35, 0.40, 0.45, 0.50 fm are added to the original results of each isotopic chain, respectively. The three vertical gray lines denote $N=20, 28, 50$.}
    \label{Rch}
\end{figure}

The charge radius of a nucleus is a key observable that can directly reflect important features of nuclear structure, such as the emergence of neutron halos~\cite{Nortershauser:2008vp,Geithner:2008zz}, the occurrence of new magic numbers or disappearance of traditional magic numbers~\cite{ANGELI201369,Li:2021fmk}. 
In Fig.~\ref{Rch}, the charge radii of Ca, Sc, Ti, V, Cr, Mn, Fe, Co, Ni, Cu, and Zn isotopes as a function of the neutron number predicted by the DRHBc theory are compared with available data~\cite{ANGELI201369,Li:2021fmk,Sommer:2022sok,PhysRevLett.131.102501}. 
Overall, the predicted charge radii are consistent with the available data.  
We note that the apparent kinks at $N=28$ manifest this traditional shell closure in the Ca, Cr, Mn, Fe, and Ni isotopic chains, and the theoretical results agree well with the available data. 
Clearly, there is reason to believe that for the other six isotopic chains, 
the predicted $N=28$ shell closure should persist and we encourage experimental measurements of charge radii of the relevant nuclei. 

Close to the proton drip line, no obvious kinks at $N=20$ can be seen from our calculations, but the data for the Sc isotopes show a pronounced kink signaling the existence of a shell closure.
We note that in the DRHBc calculations, there exists a second minimum with a large prolate deformation, resulting in a much larger charge radius. Therefore such a kink might be due to deformation effects. However, that minimum is not the ground state in the DRHBc theory with several density functionals and therefore beyond-mean-field effects need to be investigated. We stress that such a kink cannot be seen in the neighboring Ca isotopes, and therefore more investigations are needed to understand this puzzling phenomenon. In addition, around $N=20$ there are some discrepancies between theory and experiment for Ca, Sc, and Ti isotopes, especially for the odd-even staggerings.
We note that the description of the evolution along the Ca, Sc, and Ti isotopic chains and the odd-even staggerings of charge radii have always been challenging for density functional theories. Such odd-even staggerings can be related to the neutron-proton pairing correlation, 
but the DRHBc theory does not explicitly take it into account. 
In Refs.~\cite{An:2020qgp,An:2021rlw}, a phenomenological correction term is introduced
to consider the neutron-proton pairing correlation. It successfully reproduced the odd-even effects in the Ca isotopic chain, and then was applied to study ten more
isotopic chains, i.e., oxygen, neon, magnesium, chromium,
nickel, germanium, zirconium, cadmium, tin, and lead. 
In Ref.~\cite{Reinhard:2017ugx} it was shown that the pairing gradient term controlled by the coupling constant $h_{\nabla}^{\xi}$ plays a crucial role in the Fayans energy
density functional. If this term is included, it would be possible to reproduce the observed odd-even staggerings of charge radii of Ca isotopes. One can study the charge radii of all the nuclei throughout the nuclear chart in the DRHBc theory taking into account various corrections, which we shall leave for future works. 

For the neutron number $N = 32$, 
the DRHBc theory reproduces well the relevant data. 
We note that the neutron numbers $N = 32$ and $34$ have been predicted to be magic in some works. For instance, the precise measurements of masses of $^{49-57}$Ca established prominent shell closures at $N=32,34$~\cite{Michimasa:2018obr}, and the energy of $2_1^{+}$ state of $^{54}$Ca confirmed the existence of the $N=34$ shell closure~\cite{Steppenbeck:2013mga}. Nonetheless, the charge radii studied here show no indications of a shell closure, consistent with the conclusions given in Refs.~\cite{Kortelainen:2021raz,GarciaRuiz:2016ohj}. We further note that the experimental charge radii of potassium isotopes do not show signs of shell closure at $N=32$~\cite{Koszorus:2020mgn} either.

As can be seen in Fig.~\ref{Rch},  the DRHBc results agree well with the available data for Zn and Cu isotopes around the $N=50$ shell closure and the kinks at $N=50$ indicate the appearance of this shell closure.
As $Z$ decreases, the kinks at $N=50$ gradually disappear, 
indicating that the $N=50$ shell closure becomes weaker and eventually disappears. 
Nonetheless, due to the limited experimental data, further investigation is required to confirm the presence or disappearance of the $N=50$ shell closure in these isotopic chains from the perspective of charge radii.

\subsection{Two-neutron separation energies}\label{Two-neutron separation energies}
\begin{figure}[!htbp]
    \centering
    \includegraphics[width=.5\textwidth]{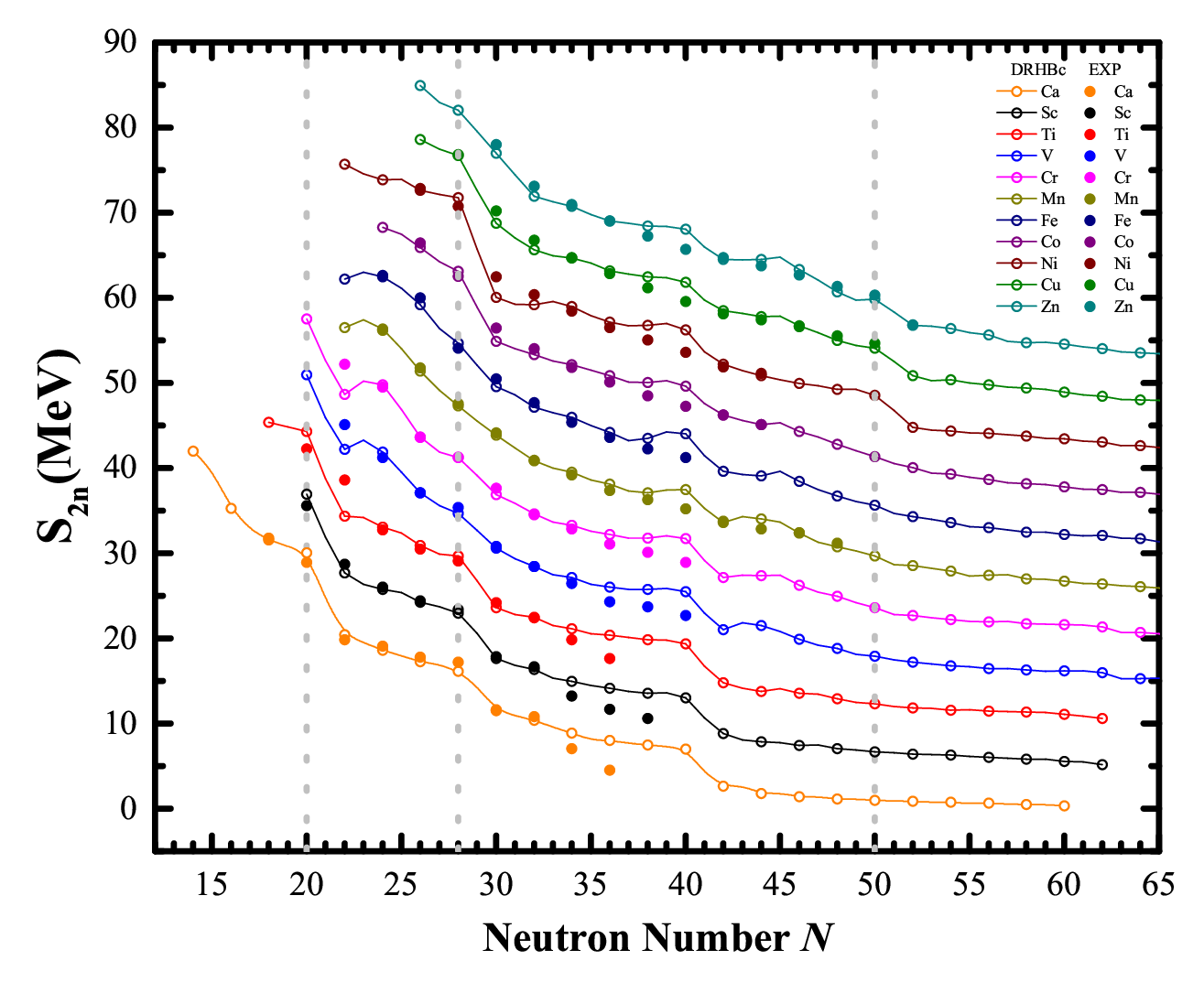}
    \caption{Theoretical two-neutron separation energies $S_{2n}$ as a function of the neutron number for Ca, Sc, Ti, V, Cr, Mn, Fe, Co, Ni, Cu, and Zn isotopes, where the results for the even-even nuclei are taken from Ref.~\cite{DRHBcMassTable:2022uhi}. The available data from Ref.~\cite{Wang_2021} are shown for comparison. For clarity, 5, 10, 15, 20, 25, 30, 35, 40, 45, 50 MeV are added to the original results of each isotopic chain, respectively. The solid points denote the experimental data and the hollow ones are the DRHBc results. The three vertical gray lines denote $N=20, 28, 50$.}
    \label{S2n}
\end{figure}

In addition to charge radii, two-neutron separation energies are also important observables to provide detailed information about shell evolution and shape transitions. 
Two-neutron separation energies $S_{2 n}$ are defined as (\ref{s2n}), 
\begin{align}\label{s2n}
 S_{2n}(Z,N)=E_{B}(Z,N)-E_{B}(Z,N-2),
\end{align}
where $E_{B}(Z,N)$ is the binding energy of a given nucleus with $Z$ protons and $N$ neutrons. 
Fig.~\ref{S2n} shows the two-neutron separation energies $S_{2n}$ as a function of the neutron number $N$ for the eleven isotopic chains studied, along with the available experimental data taken from Ref.~\cite{Wang_2021}. In general, for a given isotopic chain, $S_{2n}$ decreases smoothly with increasing neutron number $N$, except at a magic number where $S_{2n}$ drops significantly. From Fig.~\ref{S2n}, one can clearly see the $N=28$ shell closure from both the theoretical results and experimental data and that the sudden decreases of $S_{2n}$ for the Mn and Fe isotopic chains are not so obvious compared with other isotopic chains.

Moreover, near the $N=50$, the theoretical two-neutron separation energies are in good agreement with the experimental data~\cite{Wang_2021} for Cu and Zn isotopes, similar to the case of charge radii shown in Fig.~\ref{Rch}. This demonstrates that the $N=50$ shell closure in Cu and Zn is well reproduced in the DRHBc theory. Furthermore, it is worth mentioning that the disappearance of the $N=50$ shell closure for $Z=20\sim 27$ is consistent with the conclusion drawn from the charge radii studied in Sec.~\ref{charge radii}. In addition, we found that the DRHBc calculations support the appearance of the subshell closures at $N=40$ at the mean-field level for most of the isotopic chains in question but the available data do not.
This is related to the island of inversion of $N=40$ ~\cite{Sato:2012jj,Lenzi:2010zzb,Mougeot:2018ttn,Nowacki:2016isq}.
A proper theoretical description of this mass region needs to consider beyond-mean-field effects.
 
In Sec.~\ref{charge radii}, we note that the $N=20$ shell closure is not evident for Ca, Sc, Ti isotopes in our calculations. 
From the calculated two-neutron separation energies, one can still notice the sharp decreases at $N=20$ for these three isotopic chains, suggesting that this shell closure is still prominent. However, the existing data for the Ti isotopes suggest the contrary. As a result, more investigations are needed before a firm conclusion can be drawn. 

\subsection{Two-neutron gaps}
\begin{figure*}[!htbp]
    \centering
    \includegraphics[scale=0.6]{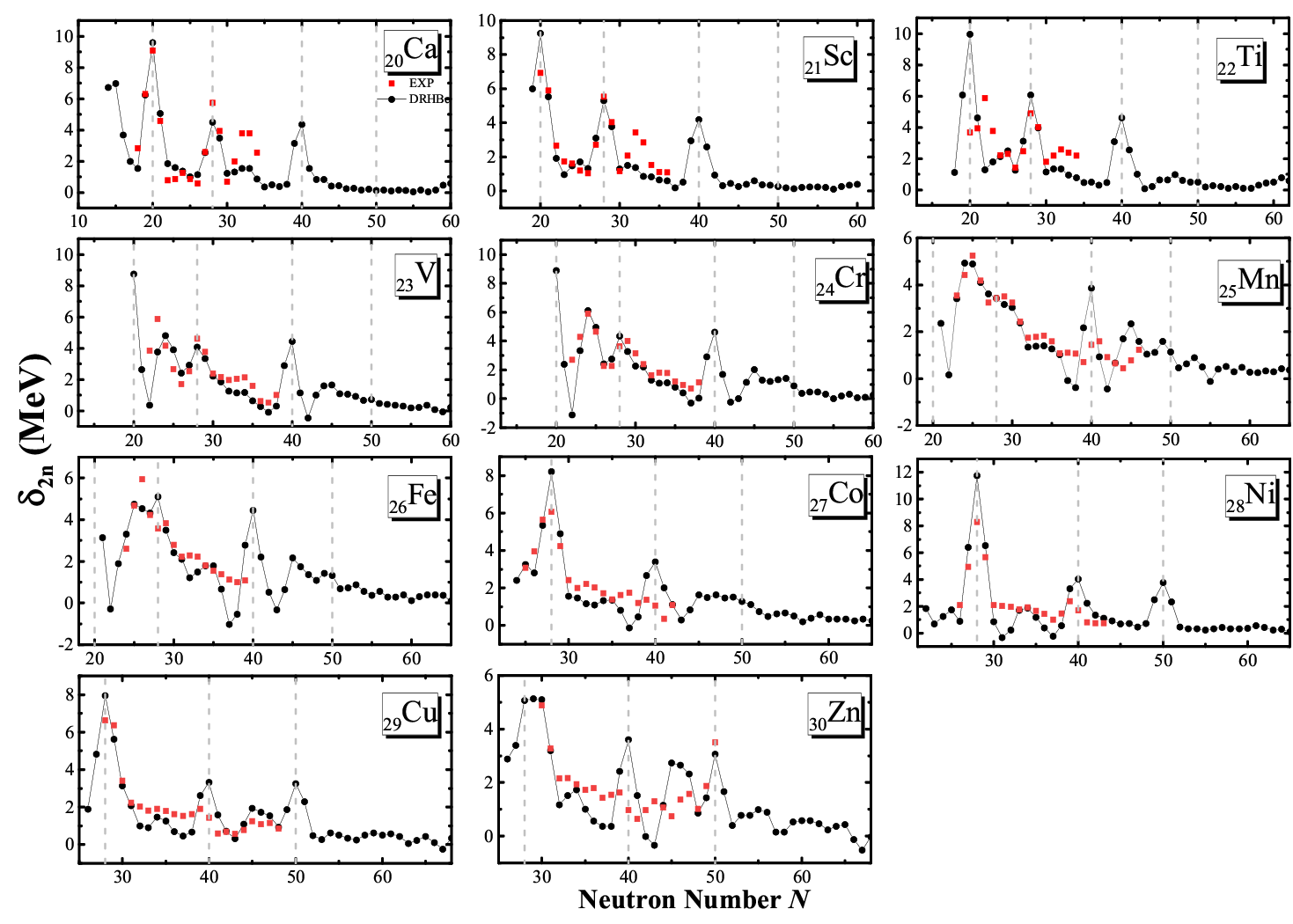}
    \caption{Theoretical two-neutron shell gaps as a function of the neutron number for Ca, Sc, Ti, V, Cr, Mn, Fe, Co, Ni, Cu, and Zn isotopes, where the results for the even-even nuclei are taken from~\cite{DRHBcMassTable:2022uhi}. The available data from Ref.~\cite{Wang_2021} are shown for comparison. The four vertical gray lines denote $N=20, 28, 40, 50$.}
    \label{deltas2n}
\end{figure*}

Compared to two-neutron separation energies, two-neutron gaps,
defined as $\delta_{2 n}=S_{2 n}(Z, N)-S_{2 n}(Z, N+2)$, are more sensitive to shell effects, because they exhibit a sharp peak when crossing a magic number (shell closure). 
The two-neutron gaps are shown in Fig.~\ref{deltas2n} as a function of the neutron number $N$ and compared with the available data~\cite{Wang_2021}, where the sharp peaks indicate the appearance of shell closures.
Clearly, the theoretical and experimental results for the $N=28$ shell closure agree well with each other, both showing a sharp peak at the traditional magic number $N=28$. 
For the Mn isotopic chain, there is no peak at $N=28$, 
indicating that the $N=28$ shell closure is quenched due to deformation effects, 
which drive this nucleus prolate in the ground state as shown in Fig.~\ref{deformation}.
In the theoretical results for the Ca, Sc, and Ti isotopic chains, the $ N=20$ shell closure is evident. We note that there is a significant discrepancy in the Ti isotopic chain between theory and experiment. The theoretical results peak at $N=20$ but the data~\cite{Wang_2021} peak at $N(=Z)=22$. This discrepancy can be attributed to the relatively strong neutron-proton pairing contributing to this particular nucleus (and the adjacent ones)~\cite{Sandulescu:2015vva,Frauendorf:2014mja}, which can distort the relation between two-neutron gaps and shell closures. Indeed, as shown in Fig.~\ref{N=20}, the neutron single-particle levels of $^{42}$Ti show that the $N=20$ shell closure is still prominent.

Regarding the $N=32$ shell closure, the experimental data show a sharp peak in the Ca, Sc, and Ti isotopic chains, while the DRHBc results only exhibit small fluctuations.  In Refs.~\cite{Steppenbeck:2013mga,Otsuka:2022zel}, it was found that the $N=32$ subshell closure is a direct consequence of the weakening of the attractive nucleon-nucleon interaction between protons $(\pi)$ and neutrons $(v)$ in the $\pi f_{7 / 2}$ and $v f_{5 / 2}$ single-particle orbitals (SPOs) as the number of protons in the $\pi f_{7 / 2}$ SPOs decreases and the magnitude of the $\pi f_{7 / 2}-v f_{5 / 2}$ energy gap increases. Therefore, for the $N=32$ shell closure, the deviation of the DRHBc results from the available data can be attributed to the missing of tensor force. We note that the localized exchange terms in the PCF-PK1 density functional theory can describe well the binding energies of Ca isotopes and the discrepancies between the theoretical results and the experimental data are less than 2 MeV~\cite{Zhao:2022xhq}. 
Recently, the new magic numbers $N=32$ and 34 in the Ca isotopes have been studied in Ref.~\cite{Liu:2019yov} using the relativistic Hartree-Fock theory~\cite{Long:2007dw}, in which the tensor force is included and it was shown that the strong couplings of the Dirac inversion partners (DIPs)  of the $(\pi,\nu) {s}_{1/2}$ and $v2 p_{1 / 2}$ states play an important role in forming the subshell closures at $N = 32,34$.

As for the $N=40$ shell closure, the high-lying $2^{+}$ state observed in ${ }^{68} \mathrm{Ni}$ and its low $B\left(E 2 ; 2^{+} \rightarrow 0^{+}\right)$ value are attributed to the relatively large energy gap separating the $p f$ and $g_{9 / 2}$ orbitals~\cite{Sorlin:2002zz}. Except for the Ni isotopic chain, there is no clear shell closure evidence from the available experimental data on two-neutron separation energies $S_{2n}$ and two-neutron gaps $\delta_{2 n}$. As for the theoretical results, similar to the conclusions drawn from the two-neutron separation energies, the two-neutron gaps also suggest the existence of this subshell closure.

Furthermore, as shown in Fig.~\ref{deltas2n}, except for the Ni, Cu, and Zn isotopes, the DRHBc results do not indicate any clear shell closure at $N=50$, which is consistent with the charge radii and two-neutron separation energies shown in Fig.~\ref{Rch} and Fig.~\ref{S2n}. These results suggest that the $N=50$ shell closure disappears in the $ 20 \leqslant Z \leqslant 27$ isotopic chains, which should be checked by future experiments.

\subsection{Quadrupole deformations}
\begin{figure*}[!htbp]
    \centering
    \includegraphics[scale=0.6]{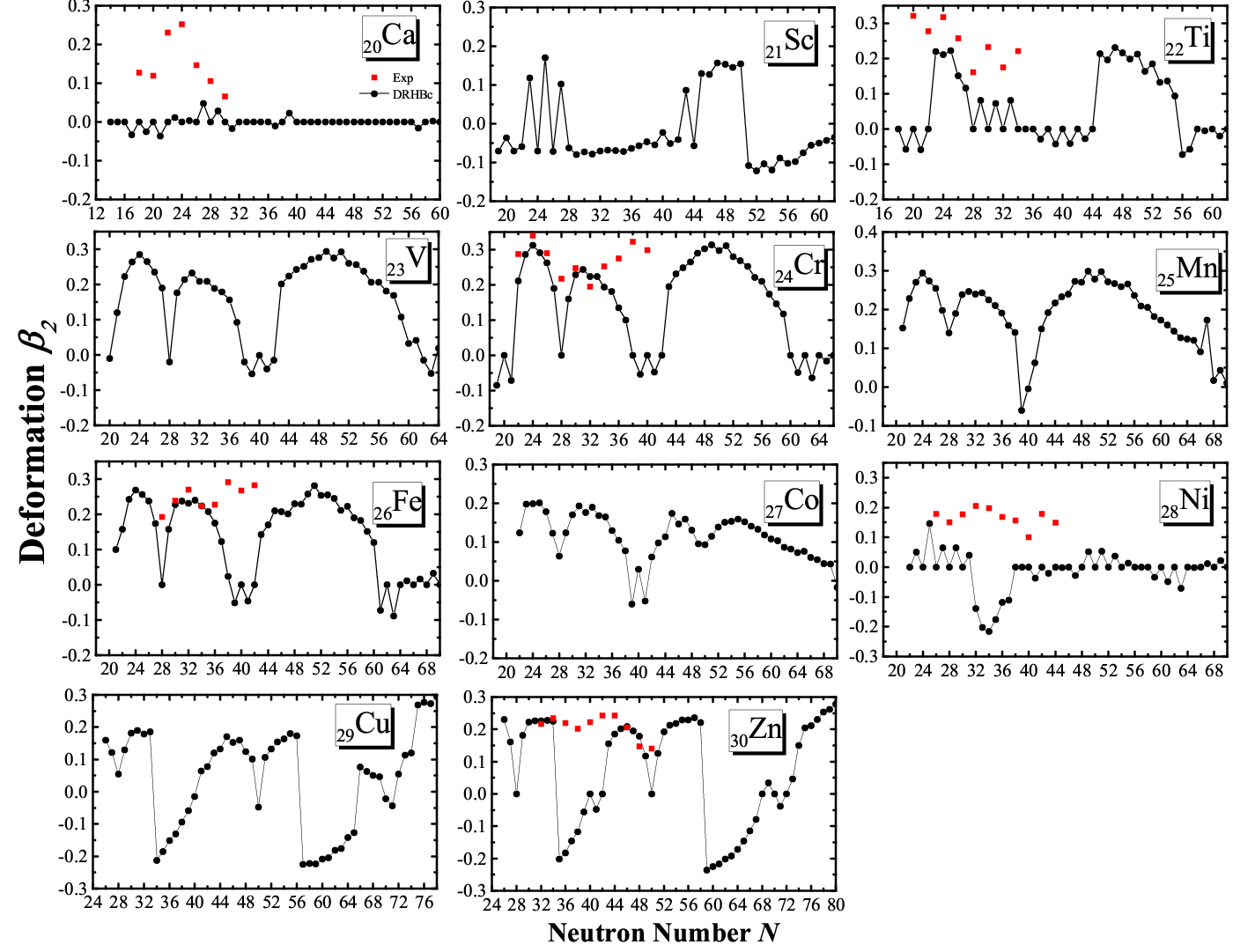}
    \caption{Theoretical quadrupole deformations as a function of the neutron number for eleven isotopic chains. The available $\beta_2$ extracted from Ref.~\cite{PRITYCHENKO20161} are also shown for comparison.}
    \label{deformation}
\end{figure*}

Intrinsic deformation, a basic property of atomic nuclei, is also influenced by the nuclear shell closure~\cite{Reinhard:2022fil}. In Fig.~\ref{deformation}, the ground-state quadrupole deformations of Ca, Sc, Ti, V, Cr, Mn, Fe, Co, Ni, Cu, and Zn isotopes as a function of the neutron number $N$ obtained using the DRHBc theory are shown and the available data taken from Ref.~\cite{PRITYCHENKO20161} are also presented.

For Ca isotopes, the deformation remains small from the proton drip line to the neutron drip line, suggesting a predominantly spherical shape throughout. The Sc isotopes exhibit a particularly notable behavior characterized by sharp peaks at $N=23$, $N=25$, and $N=27$, indicating an odd-even staggering on the deformations. As shown in Fig.~\ref{Rch}, the large deformation of $^{44}$Sc and $^{46}$Sc can partially explain their enhanced charge radii. Most Ti isotopes are spherical except for those between $N=22$ and $N=28$ and those with $44\textless N \textless 56$, which are prolately deformed. 
For V isotopes, the predicted deformations clearly reflect the possible (sub-) shell closures at $N=20, 28$, and 40 while most
other nuclei have prolate shapes in their ground states. Similar conclusions might also be drawn for  Cr, Mn, Fe and Co isotopes. It is worth noting that the ground state of $^{53}$Mn is prolate. 
For Ni isotopes, except for those with $N=25$ and $31 < N < 38$, the deformation remains small from the proton drip line to the neutron drip line. It should be noted that the experimental data of $\beta_2$ are extracted from the observed $B\left(E 2,0^{+} \rightarrow 2^{+}\right)$ values such that all values are positive.

The Ni, Cu, and Zn isotopes with $N=28, 40$, and $50$ are predicted to be spherical, 
indicating shell closures at these neutron numbers. 
The $\beta_2$ values extracted from the $B\left(E 2,0^{+} \rightarrow 2^{+}\right)$ indicate large deformation for $^{62-80}$Zn, consistent with our results except for those isotopes with $36\leq N \leq 42$ and $N=50$. We note that the available deformation data suffer the assumption of the nucleus as a rigid rotor, which might not be true for all the nuclei, in particular for (nearly) spherical nuclei~\cite{ElBassem:2019sds}.

\subsection{Single-particle levels}
Besides the bulk properties,  the mean-field single-particle levels are also good or even more transparent indicators for shell closures.
In the following, to better understand the results observed in charge radii, two-neutron separation energies, two-neutron gaps, and quadrupole deformations, we study in detail the relevant single-particle levels obtained from the DRHBc theory with the PC-PK1 density functional. 
\subsubsection{$N=20$ shell closure}

\begin{figure}[!htbp]
    \centering
    \includegraphics[width=0.45\textwidth]{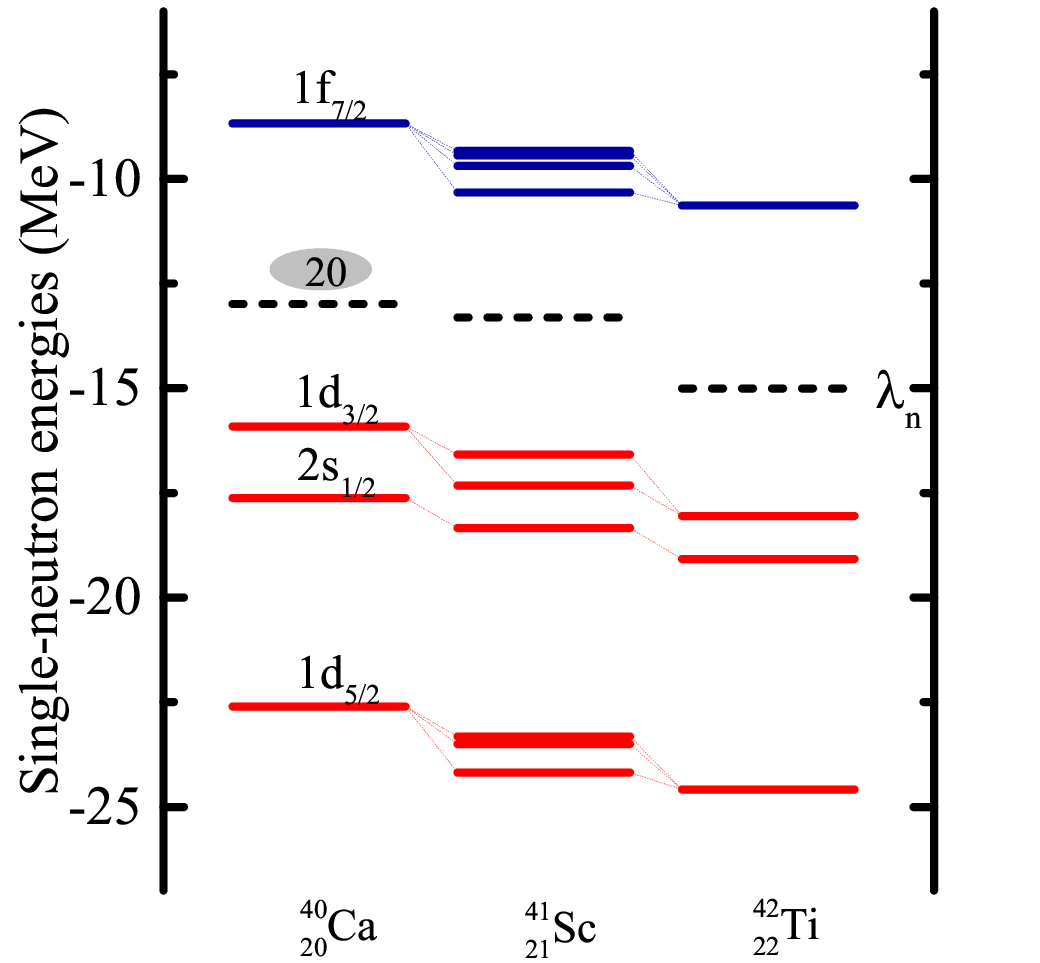}
    \caption{Theoretical single-neutron levels around the Fermi energy in the canonical basis for the ground states of $^{40}\rm{Ca}$, $^{41}\rm{Sc}$, and $^{42}\rm{Ti}$. The neutron Fermi energies ${\lambda}_{n}$ are denoted by the dotted lines, where the blue lines correspond to parity $\pi=-$ and the red lines correspond to $\pi=+$.}
    \label{N=20}
\end{figure}

To shed more light on the neutron shell closure at $N=20$, we display in Fig.~\ref{N=20} the single-neutron levels as a function of the proton number $Z$ for $20 \leq Z \leq 22$. In Fig.~\ref{deformation}, the DRHBc theory predicts that $^{40}\rm{Ca}$ and $^{42}\rm{Ti}$ are spherical, while $^{41}\rm{Sc}$ is nearly spherical with $\beta_2=-0.03$.  Fig.~\ref{N=20} shows that the $N=20$ shell closure originates from the large energy gap between the $1d_{3/2}$ and $1f_{7/2}$ orbitals, which is consistent with the Nilsson diagram ~\cite{RING1996193}.  We further note that the energy gap between the $1d_{5/2}$ and $2s_{1/2}$ orbitals is also considerably large, which might result in a subshell at $N=14$.  
  
\subsubsection{$N=28$ shell closure}
\begin{figure}[!htbp]
    \centering
    \includegraphics[width=0.48\textwidth]{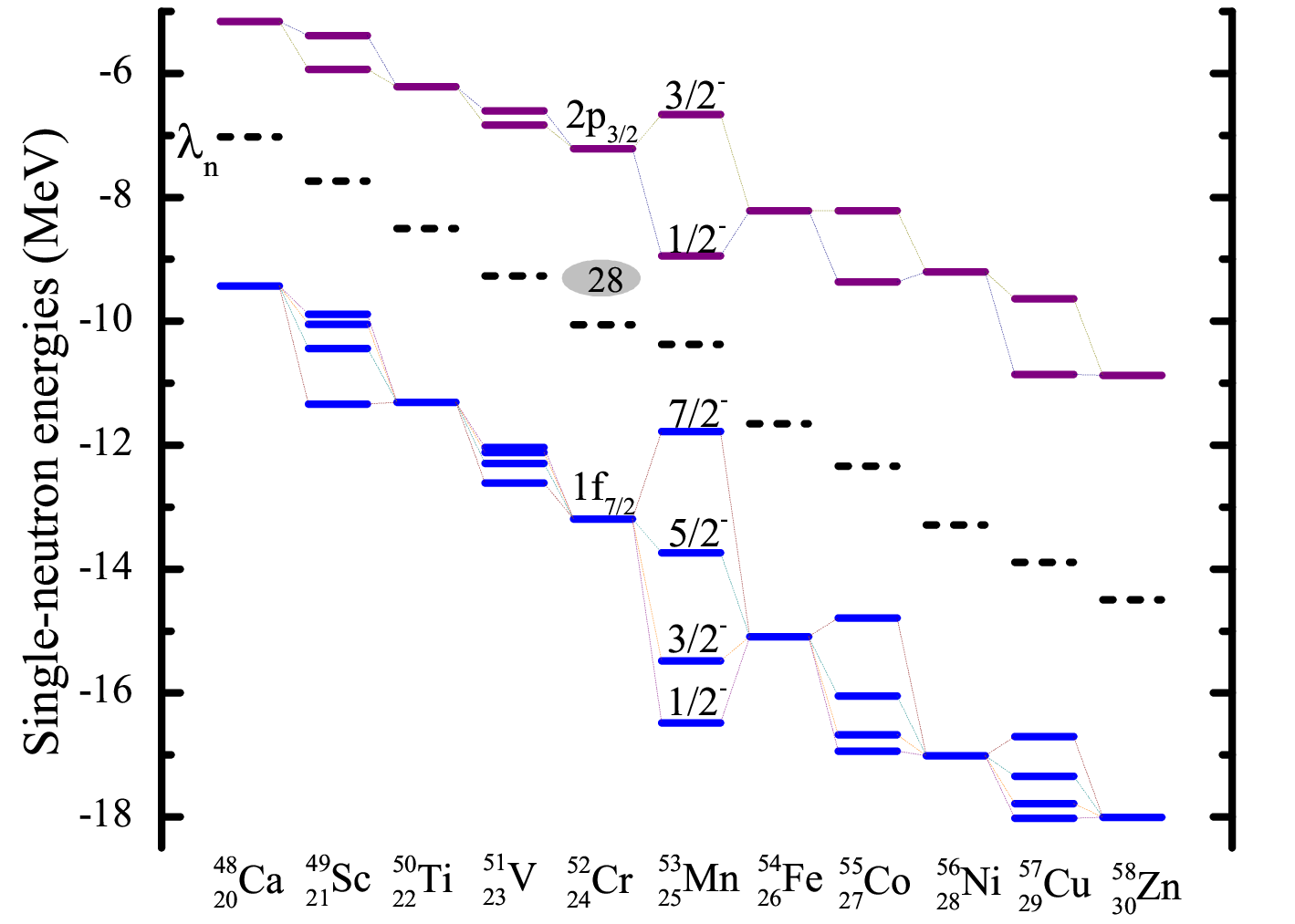}
    \caption{Theoretical single-neutron levels around the Fermi levels in the canonical basis for $N=28$ isotopes. 
    The blue lines denote $2p_{3/2}$ or those levels split from $2p_{3/2}$ for $\beta_2\ne 0$. 
    Similarly, the purple lines are for $1f_{7/2}$. }
    \label{N=28}
\end{figure}

In the following, we focus on the single-neutron levels of the $1f_{7/2}$ and $2p_{3/2}$ orbitals which are relevant to the neutron shell gap at $N=28$. In Fig.~\ref{deformation}, one can see that almost all of the $N=28$ isotopes are spherical or nearly spherical. This is consistent with the large shell gaps at $N=28$ shown in Fig.~\ref{N=28}. We note that for $^{53}$Mn whose ground state shows a prolate shape, because of deformation effects, the shell gap between the $1f_{7/2}$ and $2p_{3/2}$ orbitals, forming the $N=28$ shell closure in the spherical case, is quenched. 

\begin{figure}[!htbp]
    \centering
    \includegraphics[width=0.48\textwidth]{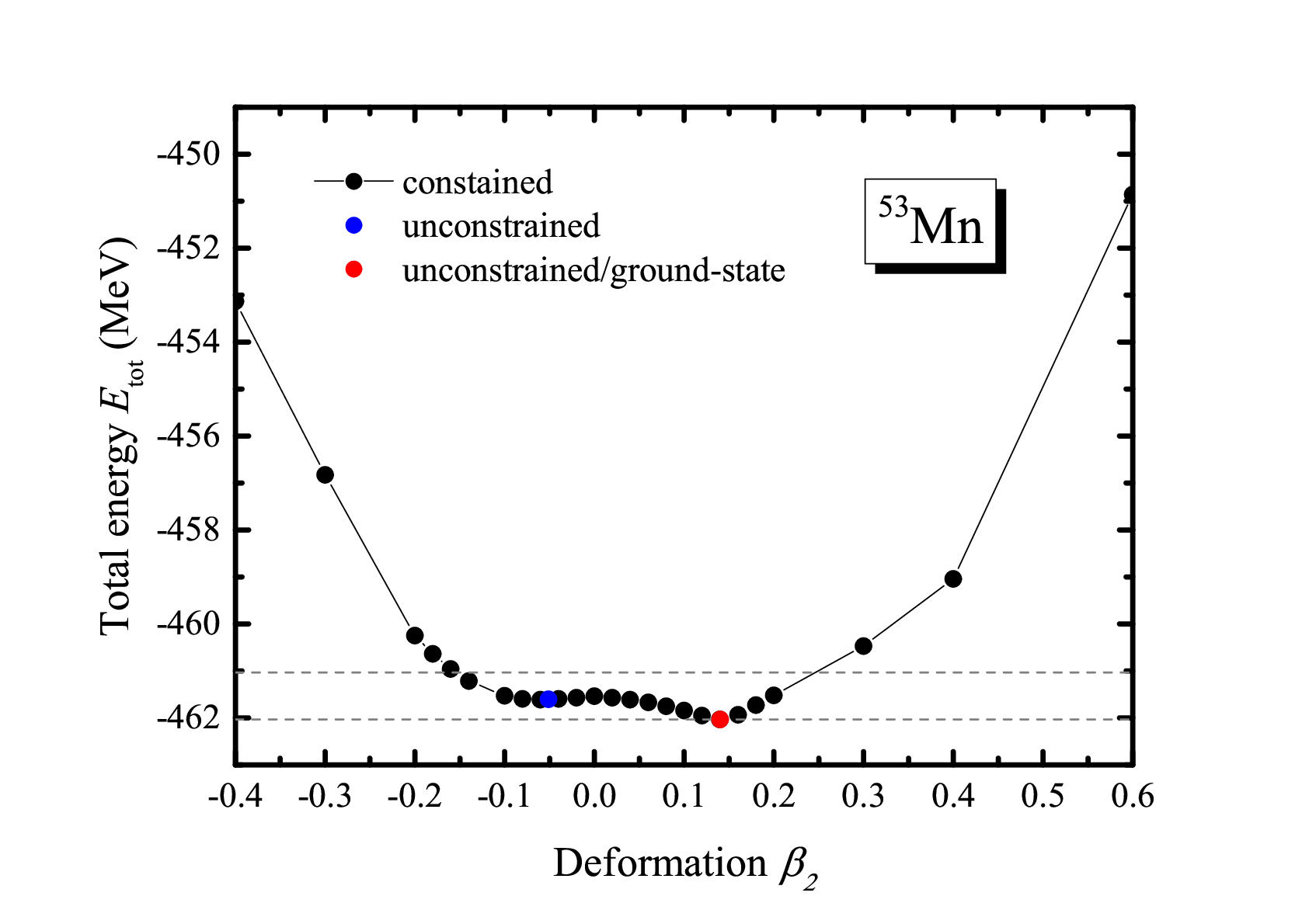}
    \caption{Potential energy curve (PEC) of $^{53}\rm{Mn}$ from the constrained DRHBc calculation. The ground state is denoted by the red solid circle. The gray lines indicate the global minimum and the energy higher by 1 MeV.}
    \label{53MnPEC}
\end{figure}

To better understand the reduction of the $^{53}\rm{Mn}$ shell gap, we show in Fig.~\ref{53MnPEC} its
potential energy curve (PEC) obtained from the constrained DRHBc calculations. 
We note that the PEC is relatively flat between $\beta_{2}= -0.1 $ and $\beta_2=0.2$.
There are two local minima: one is prolate and the other one is oblate. However, the energy difference between them is less than 0.5 MeV. The ground state has a prolate shape such that the shell closure at $N=28$ in $^{53}$Mn is quenched due to the deformation effects.  The flat PEC highlights the need for a more careful beyond-mean-field study of this particular nucleus.

\subsubsection{$N=50$ shell closure}
\begin{figure}[!htbp]
    \centering
    \includegraphics[width=0.48\textwidth]{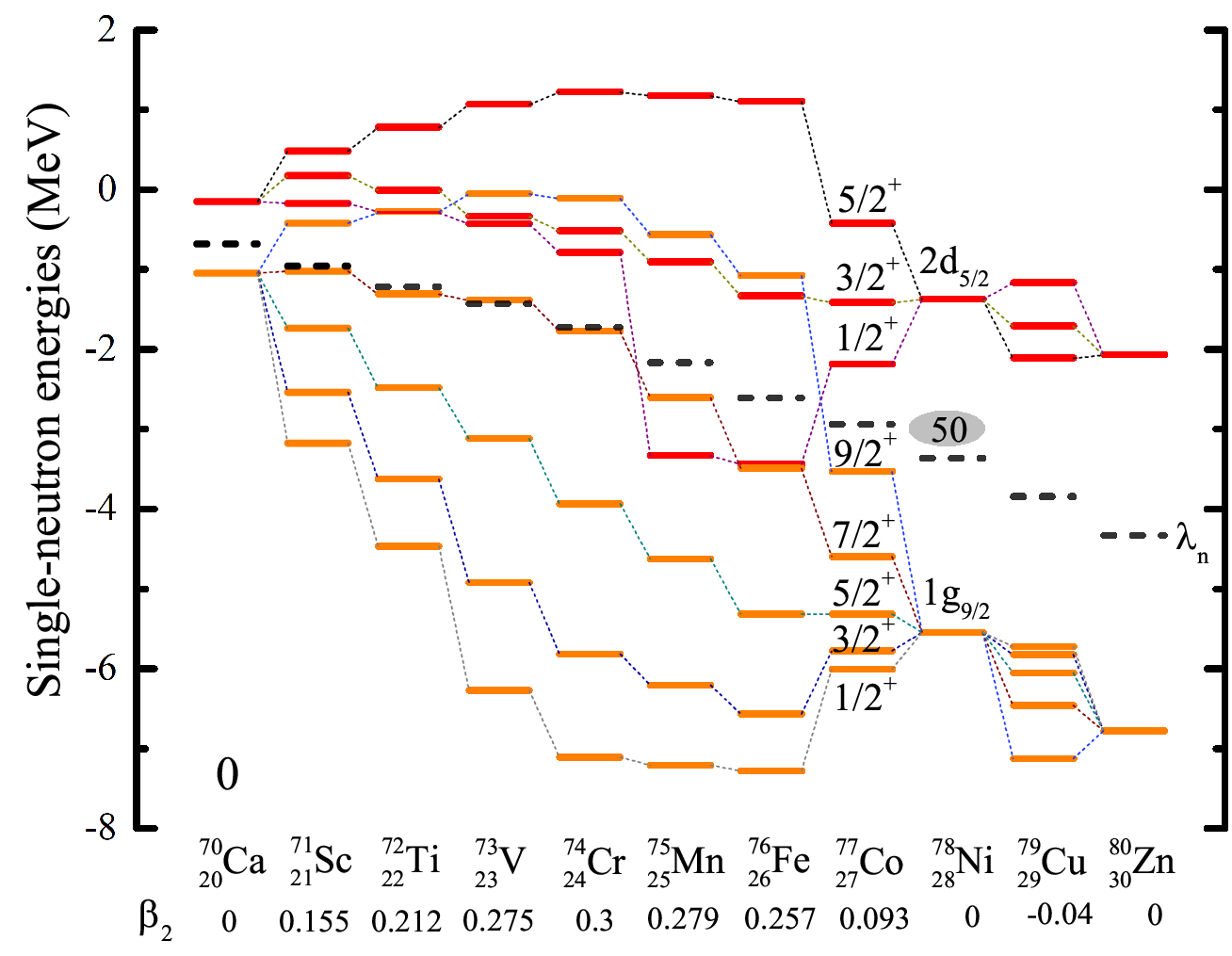}
    \caption{Theoretical single-neutron levels around the Fermi levels in the canonical basis for $^{70}\rm{Ca}$, $^{71}\rm{Sc}$, $^{72}\rm{Ti}$, $^{73}\rm{V}$, $^{74}\rm{Cr}$, $^{75}\rm{Mn}$, $^{76}\rm{Fe}$, $^{77}\rm{Co}$, $^{78}\rm{Ni}$, $^{79}\rm{Cu}$, and $^{80}\rm{Zn}$.  The neutron Fermi energy levels ${\lambda}_{n}$ are denoted by the dotted lines, where all the single-neutron levels shown have parity $\pi=+$. The predicted quadrupole deformations are indicated below each nucleus.}
    \label{N=50}
    
\end{figure}

\begin{figure}[!htbp]
    \centering
    \includegraphics[width=0.48\textwidth]{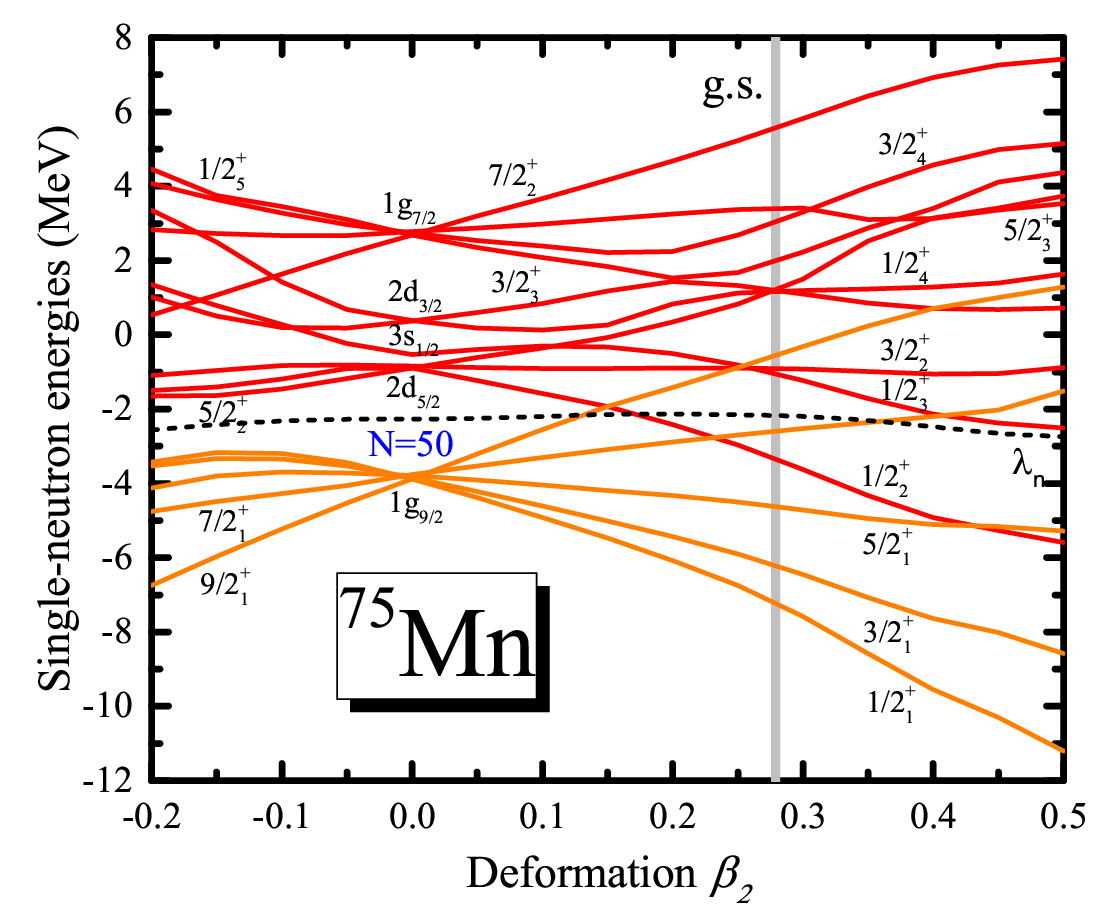}
    \caption{Theoretical single-neutron levels around the Fermi energy $\lambda_n$(dashed line) of $^{75}\rm{Mn}$ in the canonical basis from constrained calculations. The grey vertical line denotes the ground state (g.s.) of $^{75}\rm{Mn}$.}
    \label{75Mn}
\end{figure}

We now study the evolution of the $N=50$ shell closure. In Fig.~\ref{N=50}, the single-neutron levels as a function of the proton number $Z$ are shown. For  $^{78}\rm{Ni}$, $^{79}\rm{Cu}$, and $^{80}\rm{Zn}$, there is a large gap between the $1g_{9/2}$ and $2d_{5/2}$ orbitals, corresponding to the $N=50$ shell closure.  On the other hand, in Fig.~\ref{deformation}, the DRHBc theory predicts  $^{71}\rm{Sc}$, $^{72}\rm{Ti}$, $^{73}\rm{V}$, $^{74}\rm{Cr}$, $^{75}\rm{Mn}$, $^{76}\rm{Fe}$, and $^{77}\rm{Co}$ to be well-deformed, with their deformation parameters $\beta_2$ also given in Fig.~\ref{N=50}.  
The large prolate deformations of  $ ^{71}\rm{Sc}$, $^{72}\rm{Ti}$, $^{73}\rm{V}$, $^{74}\rm{Cr}$, $^{75}\rm{Mn}$, $^{76}\rm{Fe}$, and $^{77}\rm{Co}$ imply the quenching of the $N=50$ shell closure. To better understand this, taking $^{75}$Mn as an example, the evolution of the single-neutron levels around the Fermi energy with the quadrupole deformation obtained from constrained calculations is shown in Fig.~\ref{75Mn}. The ground-state deformation of $^{75}\rm{Mn}$ is indicated by the grey vertical line. In the spherical limit, there is a large energy gap between $1g_{9/2}$ and $2d_{5/2}$, forming the $N=50$ shell closure. However, strong quadrupole correlations from the mixing of $sd$ and $dg$ orbitals drive $^{75}\rm{Mn}$ prolate with $\beta_{2}=0.279$ in the ground state and increase the level density around the Fermi surface, leading to the disappearance of the $N=50$ shell closure. Therefore, one can conclude that deformation effects play an important role in the description of $^{71}\rm{Sc}$, $^{72}\rm{Ti}$, $^{73}\rm{V}$, $^{74}\rm{Cr}$, $^{75}\rm{Mn}$, $^{76}\rm{Fe}$, and $^{77}\rm{Co}$, which lead to the disappearance of certain conventional shell closures~\cite{PhysRevC.85.064329}.  

\section{Summary}\label{Summary and Outlook}
In summary, we have studied those nuclei with  $ 20 \leqslant Z \leqslant 30$ ranging from the proton drip line to the neutron drip line in the deformed relativistic Hartree-Bogoliubov theory in continuum (DRHBc) with the PC-PK1 functional. 
We have examined the evolution of the $N=20,28,50$ shell closures in this region and analyzed the charge radii, two-neutron separation energies $S_{2n}$, two-neutron shell gaps $\delta_{2n}$, quadrupole deformations $\beta_2$, and single-particle levels. Our results show that the traditional neutron shell closures at $N=20$ persist in the DRHBc theory, consistent with existing data. But the existing data for the Ti isotopes suggests the contrary, therefore more investigations are needed. It is interesting to note that the deformation effects play an important role in the description of $(21 \leq Z \leq 27)$ nuclei, which may lead to the disappearance of the $N=50$  shell closure. Similarly, both experiments and theory indicate that the $N=28$ shell closure disappears in the Mn isotopes, predominantly due to the influence of deformation effects. We encourage further investigations to verify these findings.

As a byproduct of our study of shell closures, we predicted the neutron and proton drip line nuclei of $20\le Z\le30$ for the first time with pairing, deformation, continuum, and blocking effects properly taken into account. The neutron drip line nuclei are $^{80}\rm{Ca}$, $^{83}\rm{Sc}$, $^{84}\rm{Ti}$,  $^{87}\rm{V}$, $^{90}\rm{Cr}$, $^{95}\rm{Mn}$, $^{96}\rm{Fe}$, $^{97}\rm{Co}$, $^{98}\rm{Ni}$, $^{107}\rm{Cu}$, $^{110}\rm{Zn}$, and the proton drip line nuclei are $^{34}\rm{Ca}$, $^{40}\rm{Sc}$, $^{40}\rm{Ti}$,  $^{43}\rm{V}$, $^{43}\rm{Cr}$, $^{46}\rm{Mn}$, $^{47}\rm{Fe}$, $^{49}\rm{Co}$, $^{50}\rm{Ni}$, $^{55}\rm{Cu}$, $^{56}\rm{Zn}$.

\section{Acknowledgements}
 Ru-You Zheng thanks Dr. Cong Pan and Dr. Kai-Yuan Zhang for many useful discussions. Helpful discussions with members of the DRHBc Mass Table Collaboration are highly appreciated. This work is supported in part by the National Natural Science Foundation of China (NSFC) under Grants No.11975041, and No.11961141004.
 Xiang-Xiang Sun is supported in part by 
 NSFC under Grants No. 12205308, and the Deutsche Forschungsgemeinschaft
(DFG) and NSFC through the  funds provided to the Sino-German Collaborative Research Center TRR110  ``Symmetries and the Emergence of  Structure in QCD''
(NSFC Grant No. 12070131001, DFG Project-ID 196253076).

\bibliographystyle{apsrev4-2}
\bibliography{CDFT.bib}

\end{document}